\documentclass{article}  
\usepackage{jamaica04}
\usepackage{graphicx}
\frompage{000} \topage{000}                                              
\def\rightmark{Hadronization via Coalescence}
\def\leftmark{V. Greco et al.}
 
\title{Hadronization via Coalescence} 
\authors{
{V. Greco$^1$ and C.M. Ko$^{1}$ %
}\\[2.812mm]
{\normalsize
\hspace*{-8pt}$^1$ Cyclotron Institute and Physics Department, Texas A\&M 
University, \\ 
TX-77843-3366 College Station, USA\\[0.2ex]
}}
\abstract{We review the quark coalescence model for hadronization 
in relativistic heavy ion collisions and show how it can explain 
the observed large baryon to meson ratio at intermediate transverse 
momentum and scaling of the elliptic flows of identified hadrons. 
We also show its predictions on higher-order anisotropic flows 
and discuss how quark coalescence applied to open- and 
hidden-charm mesons can give insight to charm quark interactions 
in the quark-gluon plasma and $J/\Psi$ production
in heavy ion collisions.}

 
\begin{document}
 
\maketitle
\setcounter{page}{1}

\section{Introduction}\label{intro}
To find the signals of the quark-gluon plasma and to study its
properties in relativistic heavy ion collisions, it is important to 
understand how quarks and gluons are converted to hadrons during
hadronization.  Because of its non-perturbative nature, hadronization 
has so far been treated only phenomenologically based on the 
statistical model, the duality model, or the coalescence model. 
The coalescence model, in which colored partons are combined into 
singlet hadron clusters, was first suggested as a possible mechanism 
of hadronization in studying pion correlations in relativistic heavy 
ion collisions \cite{lopez}. It was later applied to describe the  
chemical composition of the hadronic matter produced in heavy ion 
collisions at SPS and RHIC \cite{alcor}.  More recently, the quark 
coalescence model has been shown to explain not only the larger baryon 
to pion ratio at intermediate transverse momentum but also the 
scaling relation among the elliptic flows of identified hadrons 
that are observed in Au+Au collisions at the Relativistic Heavy 
Ion Collider (RHIC) \cite{voloshin,hwa,greco,greco2,fries,moln}.  
In this talk, the quark coalescence model and its applications to 
heavy ion collisions at RHIC will be reviewed. Besides particle 
spectra and elliptic flows, we also discuss results from the 
coalescence model on resonance production, higher-order $v_4$ 
flow, and charmed hadron production and flow. 
  
\section{The coalescence model} 

In the coalescence model, the transverse momentum spectrum of 
hadrons that consist of $n$ quarks is given by the overlap of the 
product of $n$ quark phase-space distribution function 
$f_{q}(x_{i},p_{i})$ with the Wigner distribution function 
$f_H(x_{1}..x_{n};p_{1}..p_{n})$ of the hadron, multiplied by 
the probability $g_H$ of forming from $n$ colored quarks 
a color neutral object with the spin and isospin of the hadron, i.e.,
\begin{equation}
\frac{dN_{H}}{d^2P_T}=g_{H}\int \prod_{i=1}^{n}\frac{p_{i}\cdot 
d\sigma _{i}d^{3}\mathbf{p}_{i}}{(2\pi)^{3}E_{i}}f_{q}(x_{i},p_{i})
f_{H}(x_{1}..x_{n};p_{1}..p_{n})\,\delta^{(2)}
\left(P_T - \sum_{i=1}^n p_{T,i}\right),
\label{coal1}
\end{equation} 
where $d\sigma$ denotes an element of a space-like hypersurface.

The $n$ quark phase space distribution is usually approximated by 
the product of the single quark distribution function, which 
consists of both a thermal and a minijet component, separated by 
the transverse momentum $p_0\sim 2$ GeV. For heavy ion collisions
at RHIC, the quark transverse momentum spectrum below $p_0$ is taken 
to be thermal with a temperature $T=170$ MeV and a radial flow 
velocity $\beta=0.5$, which are consistent with both experimental data 
and predictions from hydrodynamical calculations. The masses of 
thermal quarks are taken to be those of constituent quarks, 
i.e., $m_{u,d}=300$ MeV and $m_s=475$ MeV. Above $p_0$, partons 
are from the quenched pQCD minijets with power-law like transverse 
momentum spectrum \cite{gyul}, and their masses are those of 
current quarks. Both soft thermal and hard minijet partons are assumed 
to distribute uniformly in a fireball of transverse radius of 8.15 fm 
and longitudinal length of 4.35 fm, corresponding to a volume of 900 
fm$^3$. The number of quarks are then fixed by the measured 
transverse energy of about 700 GeV for Au+Au collisions 
at $\sqrt{s_{NN}}=200$ GeV.

For the Wigner distribution functions of hadrons, they are taken 
to be a sphere in both space and momentum with radii $\Delta_r$ and 
$\Delta_p$, respectively, which are further related by 
$\Delta_r\cdot\Delta_p^{-1}=1$ according to the uncertainty principle.
A good description of both pion and proton spectra can be obtained with 
a radius parameter of $\Delta_p= 0.24$ GeV for mesons and $0.35$ GeV 
for baryons.

\section{Particle spectra and elliptic flows}\label{details}

\begin{figure}[ht]
\vspace*{0cm}
\includegraphics[height=1.8in,width=2.4in,angle=0]{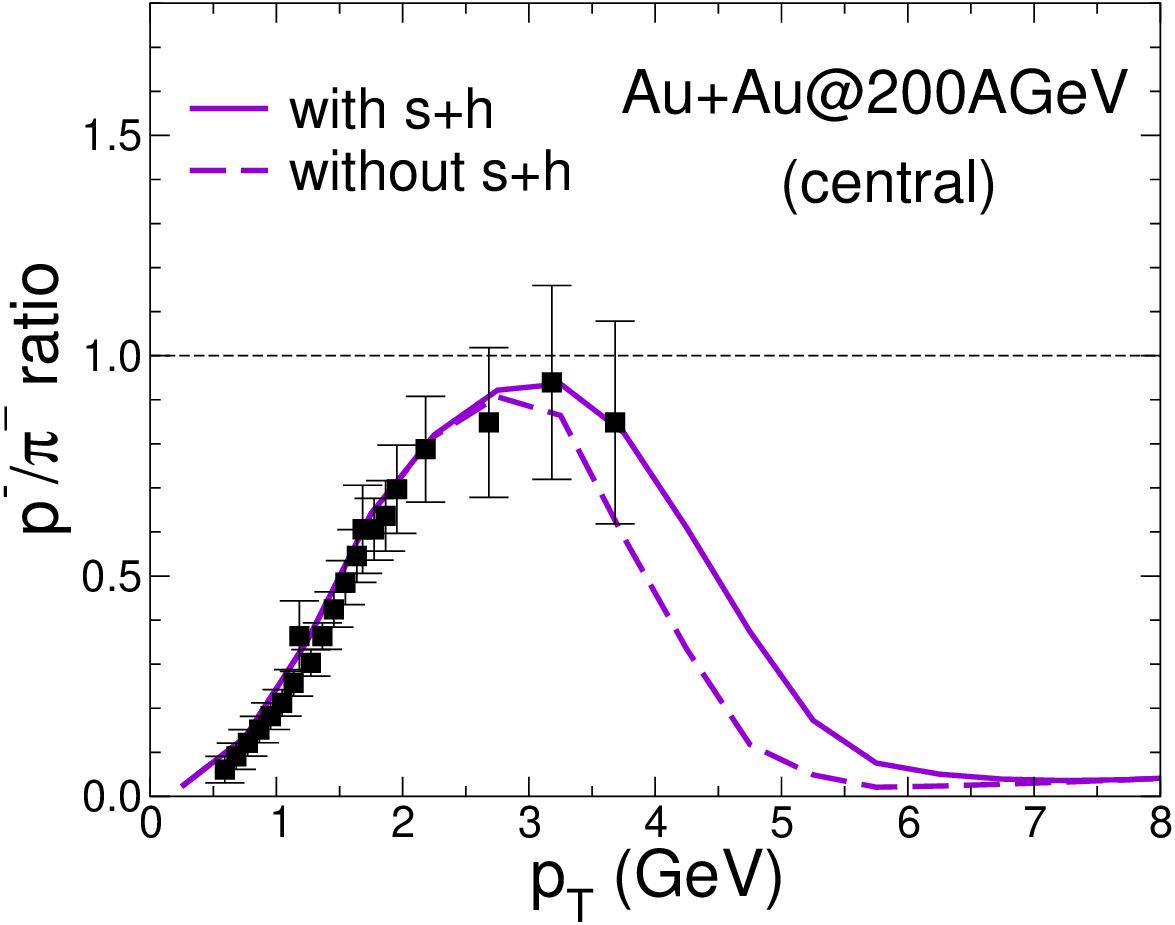}
\includegraphics[height=1.8in,width=2.4in,angle=0]{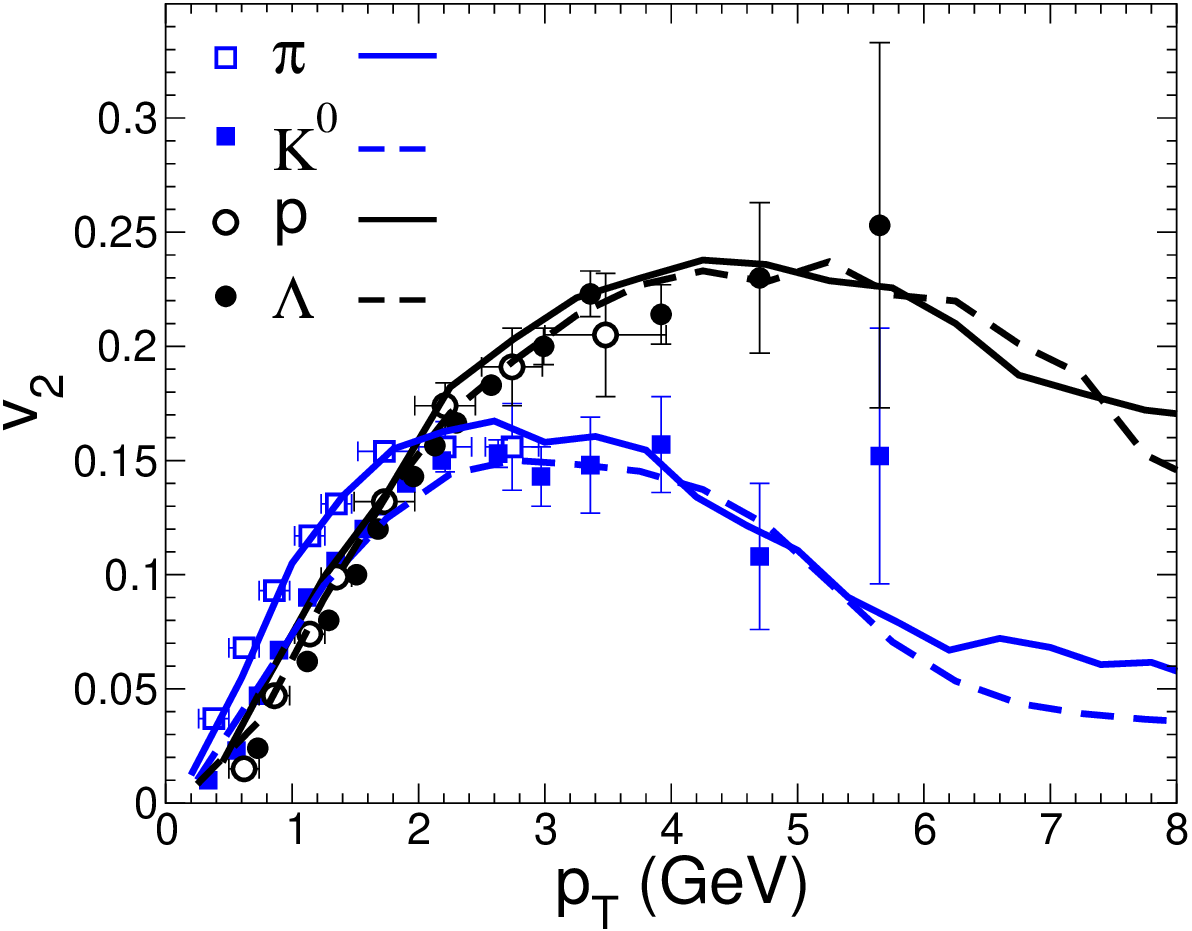}
\vspace*{-0.6cm}
\caption[]{Left panel: Antiproton to pion ratio in Au+Au collisions at
$\sqrt{s_{NN}}$=200 GeV. Solid and dashed curves are results with 
and without contribution to antiproton from coalescence of thermal 
with minijet partons. Filled squares are experimental data from 
PHENIX \cite{esumi}. Right panel: Elliptic flows of identified hadrons. 
Lines are from the coalescence model, while symbols are data from 
STAR \cite{sore} and PHENIX \cite{velk}.}
\label{fig1}
\end{figure}

In coalescence model, hadrons are formed from quarks that are close 
in phase space. As a result, baryons with momentum $p_T$ are produced 
from quarks with momenta $\sim p_T/3$, while mesons with same 
momentum are from quarks with momenta $\sim p_T/2$. Since the
transverse momentum spectra of quarks decrease with $p_T$,
production of high momentum baryons from quark coalescence is 
more favored than mesons. This is opposite to the fragmentation
process where baryon production is penalized with respect to  
mesons as more quarks need to be produced from the vacuum, leading 
to a typical $p/\pi$ ratio of $\sim 0.2$. Results on $\bar p/\pi$
ratio based on Eq.(\ref{coal1}) are shown in Fig.\ref{fig1}(left) 
together with data from PHENIX \cite{esumi}. Although different
methods have been used in evaluating the coalescence integral in
Eq.(\ref{coal1}) (for a review see \cite{fries-qm}), they all 
lead to enhanced $\bar p/\pi$ ratio (and similarly the $K/\Lambda$ ratio).
As shown in Fig.\ref{fig1}(left), our approach \cite{greco,greco2},
which includes resonance decays (to be described in the next 
paragraph) and avoids the collinear approximation by using the 
Monte Carlo method to evaluate the multi-dimensional coalescence 
integral, also gives a good description of the $\bar p/\pi$ ratio 
at $p_T < 2$ GeV. 

Another typical feature of quark coalescence is the scaling of
hadron elliptic flows according to the number of constituent quarks
in a hadron:
$v_{2,H}(p_T)/n=v_{2,q}(p_T/n)$,
which can be derived under the approximation that only quarks with
equal momentum can coalescence into hadrons \cite{moln}. Without 
any constraint on the relative momentum of coalescing quarks,
the scaling relation between baryon and meson elliptic flows
can be badly violated \cite{lin03}. In our coalescence model, we 
have taken into account the finite momentum distribution of quarks 
inside hadrons, and this leads to only a small breaking ($\sim 5\%$) 
of the so-called ``coalescence scaling'' \cite{greco2}, which is 
seen to hold within the errors of experimental data \cite{sore,velk}. 

In Fig.\ref{fig1} (right), we show how the results from our coalescence 
model can reproduce the available experimental data \cite{sore,velk} 
for $\pi,K,p,\Lambda$ once the quark elliptic flow is extracted from
a fit to the pion data \cite{greco2}. We note that the contribution of 
hadrons from minijet fragmentation is not included in this calculation. 
Its inclusion would lead to a universal hadron elliptic flow 
at momentum above $p_T \sim 6$ GeV \cite{fries}. 
In Ref.\cite{greco2}, the quark elliptic flow was extracted from 
a fit to the pion data, and its shape, especially the saturation
at $p_T \sim 1$ GeV, is in agreement with that from parton cascade 
calculations \cite{chen}. At low momenta, the elliptic flow from 
the hydrodynamical model at the phase transition temperature could
also be used in the coalescence model calculations for hadron
elliptic flows. This is because the description 
of elliptic flows at low $p_T$ are similar in the 
hydrodynamical and the coalescence model, once we correct for 
energy conservation in quark coalescence and include the residual 
radial flow effect from the hadronic stage. 


Another possible source of scaling breaking in hadron elliptic flows 
is the feed-down from resonance decays. We have studied how elliptic 
flow of stable hadrons are affected by contributions from the decay 
of resonances \cite{greco-res}.
Particles from these decays have in principle a different elliptic 
flow from that of resonances, and its value depends on the competition
between a shift in their momenta, which gives a larger elliptic flow,
and the randomization of their momenta, that decreases their initial 
azimuthal anisotropy. It turns out that particles like $p$, $\Lambda$, 
and $K$ from resonance decays have elliptic flows that are very 
similar to the directly produced ones. Therefore, the inclusion of 
resonance effect does not destroy the coalescence scaling of these 
stable hadrons. On the other hand, pions from the decay of $\omega$, 
$K^{\star}$, $\Delta$ show a significant enhancement of their elliptic 
flow at $p_T < 2$ GeV. Although, the effect of these resonances is 
reduced by pions from rho meson decays, which have an elliptic flow 
more close to that of direct pions, the overall effect of resonance 
decays on pion elliptic flow is still non-negligible. The breaking of 
coalescence scaling due to resonance decays together with that
due to finite quark momentum spread mentioned above lead to a better 
agreement with available data as shown in Fig.\ref{fig1}(left), 
where the decay of resonances was included 
\cite{greco2,greco-res}. However, one should be aware that 
such an agreement may be fortuitous especially at very low $p_T$, 
again because of the lack of energy conservation in quark coalescence.

\begin{figure}[ht]
\vspace*{-.cm}
\includegraphics[height=1.8in,width=2.35in,angle=0]{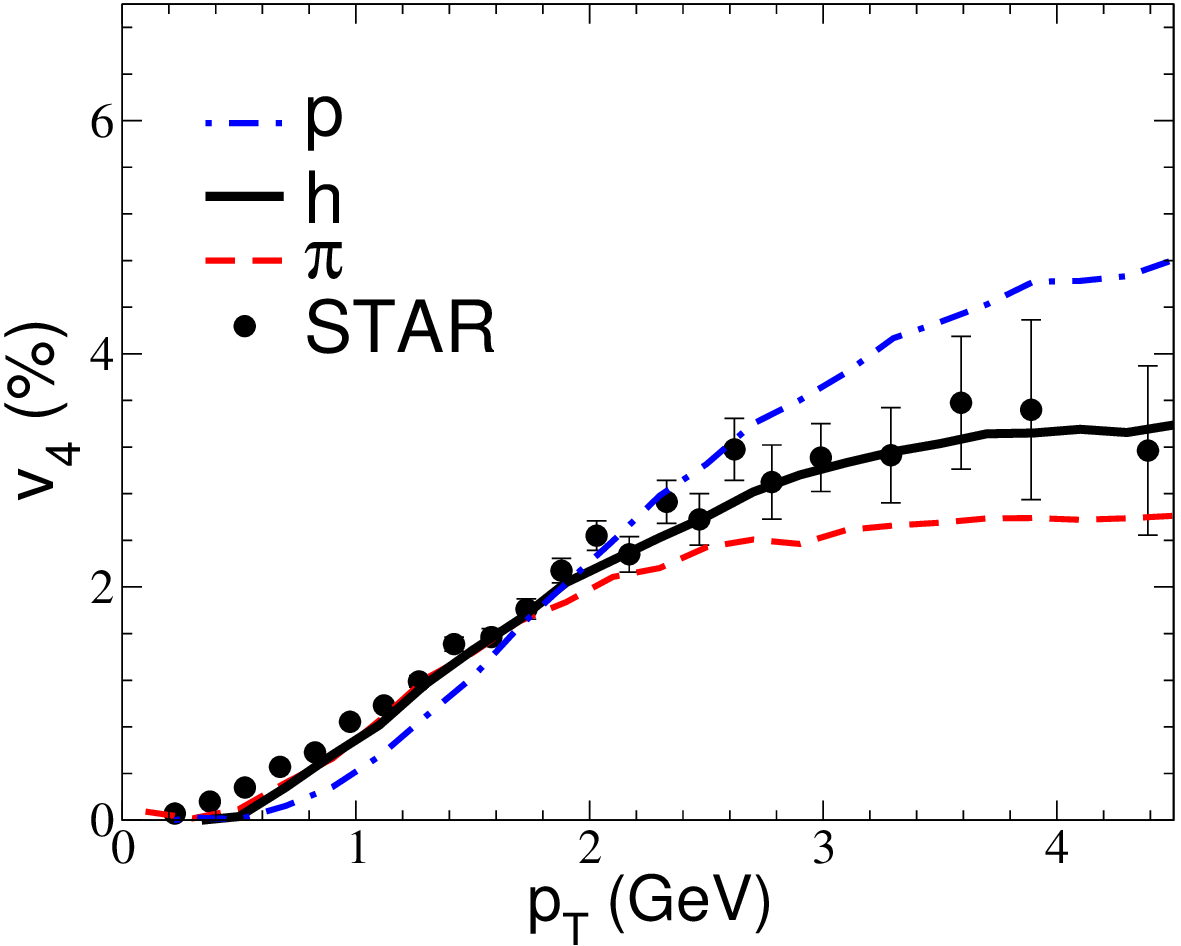}
\includegraphics[height=1.8in,width=2.45in,angle=0]{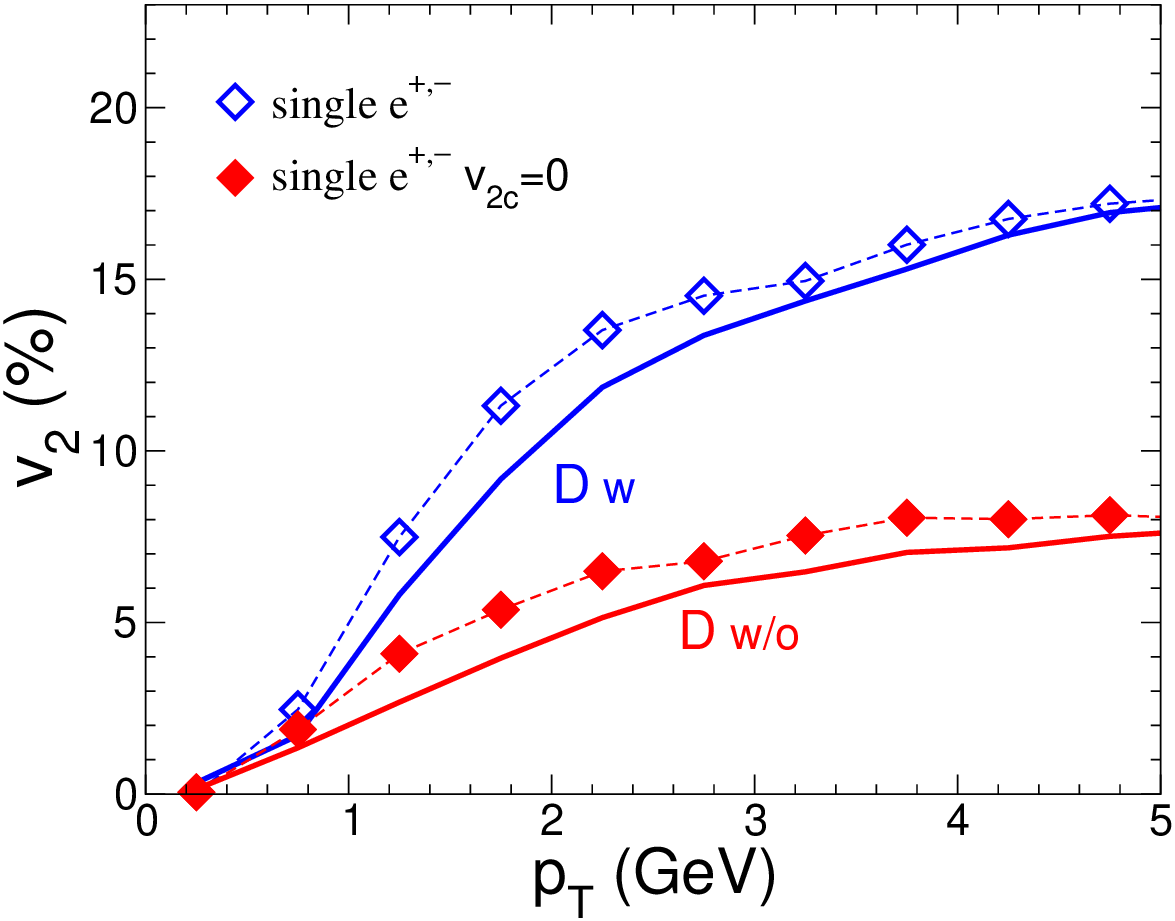}
\vspace*{-0.6cm}
\caption[]{(Left) $v_4$ for charged hadrons, pion and proton from 
quark coalescence together with the experimental data (STAR) for 
charged hadrons \cite{adams}. (Right) Elliptic flow for D mesons and 
single electrons from their decays.}
\label{fig2}
\end{figure}

Higher-order momentum anisotropy have also be measured recently
in experiments. In particular, the fourth-order harmonic $v_4$ 
was found to be sizeable \cite{adams}, as first suggested in 
Ref.\cite{kolb-v4}. Higher-order flows provide the possibility 
to further test the coalescence picture for hadronization. 
Based on the naive coalescence model that only allows quarks of equal
momentum to form a hadron, it has been shown that the meson
$v_{4,M}$ and baryon $v_{4,B}$ are related to quark $v_{2,q}$ and 
$v_{4,q}$ by \cite{coal-v4}:
\begin{equation}
v_{4,M}(2p_T)\simeq 2v_{4,q}+ v_{2,q}^2 ~\,~ v_{4,B}(3p_T)
\simeq 3v_{4,q}+3v_{2,q}^2,
\end{equation}
with $v_{n,q}$'s evaluated at $p_T$. We can see that the ratio 
between baryon and meson $v_4$ is 3 if there is no $v_{4,q}$ 
at quark level. However, a fit to the charged particle $v_4$ 
data using the quark coalescence model requires that 
$v_{4,q}\simeq 1.8 v^2_{2,q}$, which is similar to 
$v_{4,q}\sim v^2_{2,q}$ from the parton cascade \cite{chen}.
With non-zero $v_{4,q}$, $v_{4,B}/v_{4,M}$ is thus expected to be less
than 3. These results are shown in Fig.\ref{fig2} (left) together 
with experimental data on $v_4$ of charged hadrons.

\section{J/$\Psi$ and D meson}

The coalescence model can also be used to study open and hidden charmed 
hadron production in relativistic heavy ion collisions. With the large 
heavy quark mass, a more rigorous treatment of the coalescence process 
can be achieved \cite{braa}. Study of charm meson production at RHIC 
using the quark coalescence model allows us to address the following 
issues: (a) What is the sensitivity of the $J/\Psi$ abundance and its 
$p_T$ spectrum to the momentum distribution of charm quarks?  (b) How 
does the interplay of charm- and light-quark distributions translate 
into the elliptic flow of $D$-mesons? Without assuming complete 
thermalization of charm quark, which is common to statistical models, 
makes it possible to discriminate the relative importance
between $J/\psi$ regeneration and suppression in relativistic heavy 
ion collisions. Studies based on the rate equation have shown that  
regeneration of $J/\Psi$ can bring its abundance back to the 
initial value expected from the superposition of nucleon-nucleon 
interactions \cite{GR01}.  On the other hand, the initial momentum 
distribution of charm quarks from the PYTHIA routines is much 
broader than if they are thermalized in the quark-gluon plasma.
The average relative momentum between two quarks is thus larger 
than that of the charm quarks inside the charmonium. Quantitatively, 
this causes a decrease of their recombination probability 
into $J/\Psi$ by about a factor 3 with respect to the thermal 
case \cite{greco-c}.

Another way of probing charm interactions in the medium is 
through the elliptic flow of single electrons from D meson decays.
We have already shown how quark coalescence is able to describe the 
elliptic flow of light hadrons. Our knowledge of $v_2(p_T)$  
for light hadrons thus allows for a quantitative prediction of $v_{2D}(p_T)$ 
for the case in which charm quarks do not experience any final-state 
interactions. This represents the lower limit for the $v_{2D}$ of 
D mesons (D w/o in Fig.\ref{fig2} (right)), because it is only  
due to the elliptic flow of light quarks. An upper limit can be
obtained by assuming that charm quarks have a $v_{2D}$ equal 
to that of light quark (D w in Fig.\ref{fig2}(right)). It is 
also seen in Fig.\ref{fig2}(right) that single electrons 
(diamonds) from D meson decays have essentially the same flow as
that of D mesons.

In our calculations, we have not included D mesons produced from 
jet fragmentation. The elliptic flow of these D mesons depends on the
energy loss of charm quarks in the quark-gluon plasma. Since the latter
is predicted to be smaller than that of light quarks, a quite 
small $v_2$ (a first estimation put the upper limit around 5$\%$ 
\cite{djord}) is expected from these D mesons.  Experimental 
measurement of $v_{2D}$ thus allows to distinguish between the 
quark coalescence and the jet fragmentation mechanism for D meson 
production in relativistic heavy ion collisions.
  

\section*{Acknowledgments}
We thanks our collaborators, L.W. Chen, P. L\'evai, P. Kolb, and R.Rapp,  
during different phases of the presented work. This talk was based on work 
supported in part by the US National Science Foundation under Grant 
No. PHY-0098805 and the Welch Foundation under Grant No. A-1358. 
VG was further supported by by the National Institute of Nuclear 
Physics (INFN) in Italy.

\end{document}